**Title**

- Efficient light-emitting diodes based on oriented perovskite nanoplatelets

**Authors**

Jieyuan Cui[1]#, Yang Liu[2]#, Yunzhou Deng[1]#, Chen Lin[3]#, Zhishan Fang[3], Chensheng Xiang[4], Peng Bai[5], Kai Du[4], Xiaobing Zuo[6], Kaichuan Wen[7], Shaolong Gong[8], Haiping He[3], Zhizhen Ye[3]*, Yunan Gao[5], He Tian[4], Baodan Zhao[9], Jianpu Wang[7] and Yizheng Jin[1]*

**Affiliations**

[1] Zhejiang Key Laboratory for Excited-State Materials, State Key Laboratory of Silicon Materials, Department of Chemistry, Zhejiang University, Hangzhou 310027, China.

[2] Zhejiang Key Laboratory for Excited-State Materials, State Key Laboratory of Silicon Materials, School of Materials Science and Engineering, Zhejiang University, Hangzhou 310027, China.

[3] State Key Laboratory of Silicon Materials, School of Materials Science and Engineering, Zhejiang University, Hangzhou 310027, China. ZJU-WZ Novel Materials Science & Technology Innovation Center, Institute of Wenzhou, Zhejiang University, Wenzhou 325006, China.

[4] Centre of Electron Microscope, State Key Laboratory of Silicon Materials, School of Materials Science and Engineering, Zhejiang University, Hangzhou 310027, China.

[5] China State Key Laboratory for Artificial Microstructure and Mesoscopic Physics, School

of Physics, Peking University, Beijing 100871, China.

[6] X-Ray Science Division, Argonne National Laboratory, 9700 South Cass Avenue, Argonne, IL 60439, US.

[7] Key Laboratory of Flexible Electronics (KLOFE) & Institute of Advanced Materials (IAM), Jiangsu National Synergetic Innovation Centre for Advanced Materials (SICAM), Nanjing Tech University (NanjingTech), 30 South Puzhu Road, Nanjing 211816, China.

[8] Department of Chemistry, Hubei Key Lab on Organic and Polymeric Optoelectronic Materials, Wuhan University, Wuhan 430072, China.

[9] State Key Laboratory of Modern Optical Instrumentation, College of Optical Science and Engineering; International Research Center for Advanced Photonics, Zhejiang University, Hangzhou 310027, China.

* Correspondence to: Prof. Zhizhen Ye (yezz@zju.edu.cn) or Prof. Yizheng Jin (yizhengjin@zju.edu.cn)

# These authors contributed equally to this work.

**Abstract**

Solution-processed planar perovskite light-emitting diodes (LEDs) promise high-performance and cost-effective electroluminescent devices ideal for large-area display and lighting applications. Exploiting emission layers with high ratios of horizontal transition dipole moments (TDMs) is expected to boost the photon outcoupling of planar LEDs. However, LEDs based on anisotropic perovskite nanoemitters remain to be inefficient (external quantum efficiency, EQE <5%), due to the difficulties of simultaneously controlling the orientations of TDMs, achieving high photoluminescence quantum yields (PLQYs) and realizing charge balance in the films of assembled nanostructures. Here we demonstrate efficient electroluminescence (EL) from an in-situ grown perovskite film comprising of a monolayer of face-on oriented nanoplatelets. The ratio of horizontal TDMs of the perovskite nanoplatelet film is ~84%, which leads to light-outcoupling efficiency of ~31%, substantially higher than that of isotropic emitters (~23%). The nanoplatelet film shows a high PLQY of ~75%. These merits enable LEDs with a peak EQE of 23.6%, representing the most efficient planar perovskite LEDs.

**MAIN TEXT**

**Introduction**

Photon emission characteristics in semiconductors are mediated by TDMs. The optical TDMs of inorganic nanostructures with reduced dimensions, such as nanoplatelets and nanorods, are highly anisotropic(*1-4*). This unique structure-property relationship is of interest for planar LEDs because the outcoupling efficiency of the devices is fundamentally correlated to the orientation of emissive TDMs(*5-9*). In general, TDMs that are horizontally-oriented with respect to the electrode interface are favoured for light outcoupling while the vertically-oriented TDMs largely contribute to the energy loss (See fig. S1 and fig. S2 for detailed illustrations)(*10*).

Metal halide perovskite is an emerging class of solution-processed semiconductors with intriguing properties, such as high PLQYs and tunable emission wavelengths(*11-13*). Since the first report of the room-temperature-operating perovskite LEDs in 2014(*14*), remarkable progress has been made on device efficiency(*15-26*). We note that these state-of-the-art planar perovskite LEDs are based on films with isotropic TDMs (Table S1, supplementary information). Enhancing the ratio of horizontally oriented TDMs is expected to further improve light outcoupling and boost the upper limit for the EQEs of perovskite LEDs.

Here we report efficient LEDs based on in-situ grown perovskite films, which simultaneously demonstrate high ratios of horizontal TDMs and high PLQYs. The emitters are oriented perovskite nanoplatelets with a high in-plane TDM ratio of ~84%, leading to LEDs with a light extraction efficiency of ~31%. In contrast, using isotropic emitters (in-plane TDM ratio of 67%) in the same device structure would limit the light extraction efficiency to ~23%. Furthermore, we find that the PLQY of the perovskite film can be boosted to over 75% by introducing lithium bromide (LiBr) into the precursor solution. These combined efforts enable green PeLEDs with a record EQE of 23.6%, the highest value in planar PeLEDs.

**Results**

**Structural characterization of the in-situ formed nanoplatelets**

Fig. 1A shows a cross-sectional view of our device, including the perovskite layer, analyzed by aberration-corrected scanning transmission electron microscopy (STEM). Our devices consist of multilayers of nickel oxide (NiO, ∼7 nm), poly(9,9-dioctylfluorene-co-*N*-(4-butylphenyl)-diphenylamine)/poly(9-vinlycarbazole) (TFB/PVK, ∼38 nm), perovskite (~9 nm), 2,2′,2″-(1,3,5-benzinetriyl)tris(1-phenyl-1*H*-benzimidazole) (TPBi, ∼48 nm), lithium fluoride (LiF, ∼1 nm) and aluminum (Al, ∼100 nm) sequentially deposited onto indium tin oxide (ITO)-coated glass substrates (see Methods for details). The perovskite film was deposited from a precursor solution comprising of LiBr, phenylbutylammonium bromide (PBABr), phenylethyl-ammonium bromide (PEABr), caesium bromide (CsBr) and lead bromide (PbBr$_2$), dissolved in dimethyl sulfoxide (DMSO) with a molar ratio of 0.25:0.75:0.25:1.75:1.4.

The high-angle annular dark-field (HAADF) image indicates a continuous and pinhole-free perovskite film with a thickness of 8.6 ± 1.5 nm (Fig. 1A and fig. S3). Zoom-in observations (Fig. 1B) show an atomic-resolution image with well-resolved atom columns, revealing the high crystallinity of the perovskite nanoplatelets. The crystal structure of the nanoplatelet matches that of tetragonal β-CsPbBr$_3$(*27*). The perovskite crystal is oriented with the {001} crystal face parallel to the substrate surface. Atomic force microscopy characterizations on a perovskite film show a low root-mean-square surface roughness of ~1.2 nm (fig. S4), which is in line with the cross-sectional observations. The perovskite nanoplatelets were transferred and dispersed onto a copper grid for high-resolution transmission electron microscopy (HRTEM) analyses(*28*). The results (Fig. 1C and S5) indicate that the perovskite crystals are nanoplatelets with an average lateral size of 25.8 ± 6.8 nm (Fig. 1D). The Grazing-incidence wide-angle X-ray scattering (GIWAXS) measurement (Fig. 1E) of a perovskite film shows discrete diffraction spots. The diffraction spot on $q_z = 1.065$ Å$^{-1}$ corresponds to a real-space distance of 5.89 Å, which can be assigned as the d-spacing of the {001} crystal face of tetragonal β-CsPbBr$_3$. These features suggest that the assemblies of the perovskite nanoplatelets are highly ordered and all nanoplatelets share the same orientation with the {001} crystal face parallel to the substrate. The GIWAXS data, together with the STEM-HAADF and HRTEM observations, suggest that the perovskite films consist of a monolayer of face-on oriented nanoplatelets with an average thickness of 8.6 ± 1.5 nm and lateral size of 25.8 ± 6.8 nm.

**Optical analyses of the in-situ formed nanoplatelet film**

We expect the electronic and optical properties of the perovskite film to be influenced by the quantum confinement effect because the out-of-plane dimension of the nanoplatelets is comparable to the Bohr diameter of CsPbBr$_3$ (7 nm)(*29*). The Photoluminescence (PL) spectrum of the perovskite film shows the symmetric shape and its peak position is at 516 nm (Fig. 2A), corresponding to an optical bandgap of 2.41 eV. This value is larger than the bandgap of bulk β-CsPbBr$_3$, ~2.36 eV(*30*). The excitonic absorption features of quasi-two-dimensional (2D) perovskites in the strong confinement regime, such as n = 2 and n = 3 (n: the number of PbBr$_4$ octahedral layers within a crystallite) layered perovskites, are absent in the ultraviolet-visible absorption spectrum (Fig. 2A). This feature suggests that our film comprising of perovskite nanoplatelets is distinctive from the previously reported perovskite films of multiple quantum wells or quantum dots embedded in quasi-2D phase(*16, 17, 22*). Remarkably, the perovskite nanoplatelet film exhibits a high PLQY of ~75% at a low excitation power density of ~0.02 mW/cm$^2$ (Fig. 2B). This result indicates efficient radiative recombination of the photogenerated excitons in the perovskite nanoplatelets.

The orientation of TDMs of the perovskite nanoplatelet film is quantified by the ratio of horizontal TDMs, Θ. This parameter is defined as $\Theta = p_{//} / (p_{//} + p_{\perp})$, where $p_{//}$ and $p_{\perp}$ stand for contributions from horizontal and vertical components of the optical TDMs, respectively. We used the angle-dependent PL technique, which provides an ensemble measurement (photoexcitation spot size: ~1 mm) of the intensity of p-polarized emission against the detection angle (see fig. S6 for schematic illustration)(*7, 31, 32*). The experimental data (Fig. 3A) is fitted to the pattern simulated using the classical dipole radiation model(*10, 33, 34*). The Θ of our perovskite nanoplatelet film is determined to be 84%. This value is substantially higher than that of isotropic emitters (67%). Furthermore, we analysed the light emission of the perovskite film by using back focal plane (BFP) spectroscopy. BFP spectroscopy (see fig. S7 for schematic illustration) probes a small targeted region of the perovskite nanoplatelet film by employing a laser with a spot size of ~750 nm for photoexcitation(*1-3, 35*). The BFP pattern and the corresponding line-cut data along the p-polarized direction are shown in Fig. 3B and C, respectively. An important feature of the horizontally oriented dipoles is that the p-polarized intensity is minimum at $k_{//} = k_0$, where $k_{//}$ is the in-plane wavevector with respect to the substrate and $k_0$ is the wavevector in the vacuum. By fitting the line-cut data in Fig. 3C, Θ is determined to be ~87%. The BFP data of 4 spots from different regions (fig. S8) demonstrate excellent spatial uniformity of the orientations of TDMs in our film. The results of both measurements unambiguously demonstrate that the emission of our perovskite nanoplatelet film is dominated by horizontal TDMs.

We suggest that the anisotropy of local fields in the nanoplatelet, i.e., dielectric confinement, largely contributes to the anisotropy of the dipole orientation (*3, 36*). The local electric field normal to the nanoplatelet plane, $E_\perp$, is reduced (screened) due to the dielectric confinement, while the local field in the horizontal directions, $E_\parallel$, is less affected. The spontaneous radiative transition probability of the nanoplatelet, which is proportional to the square of local fields, is thus significantly reduced in the normal direction. For our closely packed oriented nanoplatelets, the local field in the horizontal directions is expected to be even less affected due to the near continuum of the high-dielectric-constant perovskite material in the lateral directions.

**Key factors affecting the perovskite films**

Two issues, namely the concentrations of the bulky organic ammonium cations and the introduction of LiBr in the precursor solution, are critical for the formation of oriented perovskite nanoplatelet film with high PLQY. Without the use of the bulky organic ammonium cations, the resulting perovskite film shows an optical bandgap of ~2.37 eV (fig. S9A) and excitation intensity-dependent PLQY (fig. S9C). These features indicate the formation of three-dimensional $CsPbBr_3$ crystals. Doubling the concentration of the bulky organic ammonium cations leads to the formation of perovskite films with strong excitonic absorption peaks at ~430 nm, ~460 nm and ~475 nm, which corresponds to the n = 2, n = 3 and n = 4 layered perovskites, respectively (fig. S9B). The emission peak of this perovskite film locates at ~501 nm, implying efficient energy transfer from the perovskites with small n values (larger bandgaps) to the emissive centers with smaller bandgaps. The emissive centers are determined to be nanocrystals with a size of ~10 nm (fig. S9D). BFP measurements on the films processed from the precursor solution with various contents of bulky organic ammonium cations indicate that only the oriented perovskite nanoplatelet films possess high ratios of horizontal TDMs (fig. S10). Furthermore, control experiments show that other investigated parameters, i.e., the concentration of the precursor solution, the choice of bulky organic ligands and the underlying substrates, do not significantly affect the $\Theta$ values of perovskite films (fig. S11). We suggest that the horizontal orientation of the perovskite nanoplatelets on flat substrates may origin from the van der Waals interactions as reported in the literature (*37, 38*). Besides, the ultra-thin film of ~9 nm would also favour horizontally aligned perovskite nanoplatelets (*39*).

Further experiments show that the introduction of LiBr in the precursor solution is beneficial for improving the PLQY of the perovskite film. GIWAXS, angle-dependent PL and BFP measurements (fig. S12) on the perovskite films processed from the precursor solution without LiBr indicate the formation of oriented nanoplatelets with anisotropic emission ($\Theta$: 84%). The PLQY of perovskite nanoplatelet films processed from the precursor solution with and without LiBr is ~75% and ~50% (fig. S13), respectively. Temperature-dependent PL characterizations (fig. S14) indicate a scenario that the with-Li sample possesses fewer energy levels below excitonic levels and thereby, fewer nonradiative losses of photo-generated excitons. We suggest that the introduction of LiBr in the precursor solution may result in perovskite nanoplatelets with better surface passivation.

**Characterization of the PeLEDs**

The EL spectrum of our perovskite nanoplatelet film (Fig. 4A) displays a symmetric peak centred at ~518 nm with a full width at half-maximum of 16 nm (74 meV), representing one of the narrowest emission line widths for high-efficiency perovskite LEDs(*20, 24-26, 40-*

*42*). The ultra-pure green emission corresponds to Commission Internationale de l'Eclairage (CIE) colour coordinates of (0.09, 0.78) (fig. S15). The angular emission intensity of our PeLEDs follows the Lambertian profile (Fig. 4B). The EL spectra at different viewing angles are identical (fig. S16). The current density−voltage−luminance (J−V−L) curves of a typical device are shown in Fig. 4C. Owing to the pinhole-free morphology of the nanoplatelet film, the device shows negligible leakage current. The current density and luminance increase rapidly once a turn-on voltage of ~3 V is reached. At 7 V, the device shows a brightness of ~3140 cd/m$^2$. The champion device demonstrates a peak EQE of 23.6% (Fig. 4D), which is a record efficiency among perovskite LEDs (Table S1). The statistical diagram of 36 devices (Fig. 4E) indicates an average EQE of 21.3% with a small relative standard deviation of 4.4%, demonstrating the excellent reproducibility of our green LEDs. The device shows a typical T$_{50}$ lifetime of ~15 min at an initial luminance of ~1080 cd/m$^2$ (fig. S17).

We performed optical simulations on the perovskite LEDs by using the classical dipole model developed for planar microcavities(*33, 34*) (see fig. S18 for details). The result suggests a light outcoupling efficiency of 31.1% for our perovskite devices based on the oriented nanoplatelet film with $\Theta$ of 84% (fig. S18B). In contrast, a control device based on isotropic TDMs ($\Theta$: 67%) would possess a lower outcoupling efficiency of ~23.4%. Considering that the PLQY of our perovskite nanoplatelet film is ~75%, our optical simulation predicts a maximum EQE of 23.3% (Fig. 4F), which agrees well with the experimental results. Further optimizing the PLQY and enhancing $\Theta$ of the perovskite films shall push the upper limit of EQEs to ~40% (Fig. 4F).

We highlight that previous work aiming to control the orientations of TDMs focuses on assemblies of anisotropic colloidal nanostructures(*43-46*). High-efficiency EL would require syntheses of anisotropic colloidal nanostructures with high PLQY, controlling the orientations of the anisotropic colloidal nanostructures to realize films with high ratios of horizontal TDMs, minimizing energy transfer between the neighbouring nanoemitters to maintain high PLQYs and efficient and balanced charge injection into the individual nanoemitters. Fulfilling these stringent requirements is challenging and often imposes dilemmas in material design and assembly. As a result, the EQEs of the LEDs based on colloidal anisotropic perovskite nanoemitters are currently lower than 5%(*47*) (see Table S2 for more information). Expanding the scope to all solution-processed planar LEDs utilizing inorganic emitters with oriented TDMs (Table S2), the reported highest EQE is 12.1%(*43*). In contrast, our LEDs based on in-situ grown perovskite nanoplatelet films demonstrate high EQEs of up to 23.6%.

**Discussion**

We demonstrate that controlling the orientation of TDMs of the perovskite films overcomes the light-outcoupling limitation of planar LEDs with isotropic emitters, leading to green LEDs with exceptionally high EQEs of up to 23.6%. Considering the chemical versatility of perovskite materials, our facile approach of in-situ grown nanoplatelet films is readily extended to the fabrication of differently coloured LEDs with high EQEs. Our work represents a simple yet effective approach of exploiting anisotropic optical properties of nanostructures in optoelectronic devices.

**Materials and Methods**

**Materials**

TFB was purchased from American Dye Source. PBA (>98.0%), PEA (>98.0%), nickel acetate tetrahydrate (>99.0%) and chlorobenzene (99.8%) were purchased from Acros.

DMSO (99.9%), caesium bromide (CsBr, 99.999%), lithium fluoride (LiF, 99.99%) and hydrobromic acid (48 wt% in water) were purchased from Alfa-Aesar. PVK (with a molecular weight of 25000−50000 g·mol$^{−1}$), lithium bromide (LiBr, 99.999%), ethanolamine (99%) and lead bromide (PbBr$_2$, 99.999%) were purchased from Sigma-Aldrich. TPBi was purchased from Xi'an Polymer Light Technology Corp.

NiO$_x$ precursor was prepared by following a literature method(*48*).

PBABr and PEABr were synthesized following the instruction of the previous report(*28*). Briefly, hydrobromic acid (2.56 mL) was slowly added into a solution of PEA (3 mL) in methanol (20 mL) at 0 °C and then stirred for 2 h. Next, the solution was evaporated at 45 °C to obtain precipitates, which were washed thrice with a mixed solution of ethyl acetate and diethyl ether with a volume ratio of 1:1. Then the precipitates were dried at 50 °C in the vacuum for 24 h.

The precursor solution for the perovskite nanoplatelet film was prepared by mixing a CsPbBr$_3$ precursor solution (1 mL), a LiBr precursor solution (0.25 mL) and DMSO (0.5 mL). The CsPbBr$_3$ precursor solution was prepared by mixing PBABr (18.9 mg), PEABr (5.5 mg), CsBr (40.8 mg), PbBr$_2$ (56.2 mg) in DMSO (1 mL), followed by stirring for 2 h at 50 °C. The LiBr precursor solution was prepared by dissolving LiBr (9.55 mg) in DMSO (1 mL).

**Device fabrication**

The NiO$_x$ precursor was deposited on ITO-coated glass substrates (square resistance of 20 Ω) by a literature method(*22*). The TFB (in chlorobenzene, 8 mg/mL) was spin-coated at 2,000 r.p.m, and then annealed at 150 °C for 30 min. Then the TFB film was spin-rinsed by chlorobenzene, leaving an ultrathin layer. Next, the PVK layer was deposited by spin-coating PVK solution (in chlorobenzene, 14 mg/mL) at 2,000 r.p.m, followed by annealing at 150 °C for 30 min. The perovskite film was prepared by spin-coating the precursor solution at 4,000 r.p.m for 2 min, followed by annealing at 100 °C for 25 min. Next, the TPBi (48 nm), LiF (1 nm), and Al (100 nm) were sequentially deposited by using a thermal evaporating system (Trovato C300) at a high vacuum environment (under $2\times10^{-7}$ Torr). The device area defined by the overlapping area of the ITO and Al electrodes was 3.24 mm$^2$.

**Structural and optical characterizations**

The cross-sectional samples were prepared by using focused ion beam (FIB) equipment (Quata 3D FEG). The TEM observations were conducted using a Cs aberration-corrected scanning transmission electron microscope Titan G2 80-200 ChemiSTEM microscope operated at 200 kV.

Atomic force microscopy analyses were conducted with a Cypher-S (Asylum Research) atomic force microscope located in a glovebox filled with nitrogen.

GIWAXS measurements were performed at Beamline 12-ID-B of Advanced Photon Source (Argonne National Laboratory, USA). A PerkinElmer detector, XRpad$^{TM}$ 4343F, was applied in the measurements with a sample-to-detector distance of 18 cm. The energy of X-ray radiation was 13.3 keV.

The angle-dependent PL spectra of the perovskite nanoplatelet films (deposited on quartz/TFB/PVK substrate) were collected by using a molecular orientation characteristic measurement system C14234-11 (Hamamatsu Photonics K. K., Japan, fig. S6). The range of angles is 0° to 90°. The excitation wavelength is 365 nm.

The BFP imaging experiments were performed on an optical system as schematically shown in fig. S7. The excitation wavelength is 457 nm and the diameter of the excitation spot is ~750 nm.

The refractive index (n) and extinction coefficient (k) of all layers of the perovskite LEDs were measured by an ellipsometer (J.A.Woollam, USA).

Ultraviolet-visible absorption spectra of the samples were collected by using a Carry 5000 (Agilent) spectrophotometer.

Steady-state PL, transient PL and temperature-dependent PL spectra were performed on an Edinburgh Instruments (FLS920) spectrometer.

Excitation-density-dependent PLQYs of the perovskite films were obtained by following the published method (*22*).

The angle-dependence of emission intensity of the perovskite LED was measured using a Thorlabs PDA100A detector at a fixed distance of 200 mm from the EL device(*16*).

All LEDs were measured in a glovebox filled with nitrogen at room temperature. A system consisting of an integration sphere (FOIS-1), a Keithley 2400 source meter and a QE-Pro spectrometer (Ocean Optics) was used for the measurements(*49*). The devices were scanned from zero bias at a rate of 0.1 $V \cdot s^{-1}$. The integral time for each step is 500 ms. The EL characteristics of the LEDs were cross-checked at Zhejiang University (Yizheng Jin group), University of Cambridge (Richard Friend group) and Nanjing Tech University (Jianpu Wang group).

**Acknowledgments**

**General**: We thank Xuan Zeng and Prof. Chuluo Yang (Wuhan University, China) for their assistance with the angle-dependent PL measurements. We thank Dr. Wei Fang (Zhejiang University, China) for the inspiring discussions on the characterizations of the optical properties of the perovskite films. This research used resources of the Advanced Photon Source, a U.S. Department of Energy (DOE) Office of Science User Facility operated for the DOE Office of Science by Argonne National Laboratory under Contract No. DE-AC02-06CH11357.

**Funding:** This work was financially supported by the National Key R&D Program of China (2016YFB0401600), the National Natural Science Foundation of China (91833303, 51911530155, 91733302 and 62005243), the Fundamental Research Funds for the Central Universities (2017XZZX001-03A and 2020QNA5002), the Key Research and Development Program of Zhejiang Province (2021C01030), the Natural Science Foundation of Zhejiang Province (LR21F050003) and China National Postdoctoral Program for Innovative Talents (BX20200288).

**Author contributions:** Y.J. and Y.L. conceived the idea and designed the experiments. Y.J. and Z.Y. supervised the work. J.C. carried out the device fabrication. J.C. and Y.L. carried out the device characterizations. Y.D. carried out the optical simulation of our perovskite devices. J.C., Y.L., C.L, and Z.F. conducted the optical measurements. P.B. and Y.G. did the BFP measurement. J.C. synthesized the PBABr and $NiO_x$. C.X., K.D., and H.T. carried out the STEM and HRTEM characterizations. X.Z. conducted the GIWAXS measurements. C.L. and Y.D. carried out the AFM experiments. B.Z. cross-checked the LED measurements. K.W. conducted the measurements of the angular dependence of EL emission. Y.J., J.C., and Y.L. wrote the first draft of the manuscript. H.H., Z.Y., Y.G. and J.W. participated in data analysis. All authors discussed the results and commented on the manuscript.

**Competing interests:** The authors declare that they have no competing interests.

**Data and materials availability:** All data needed to evaluate the conclusions in the paper are present in the paper and/or the Supplementary Materials.

**Figures and Tables**

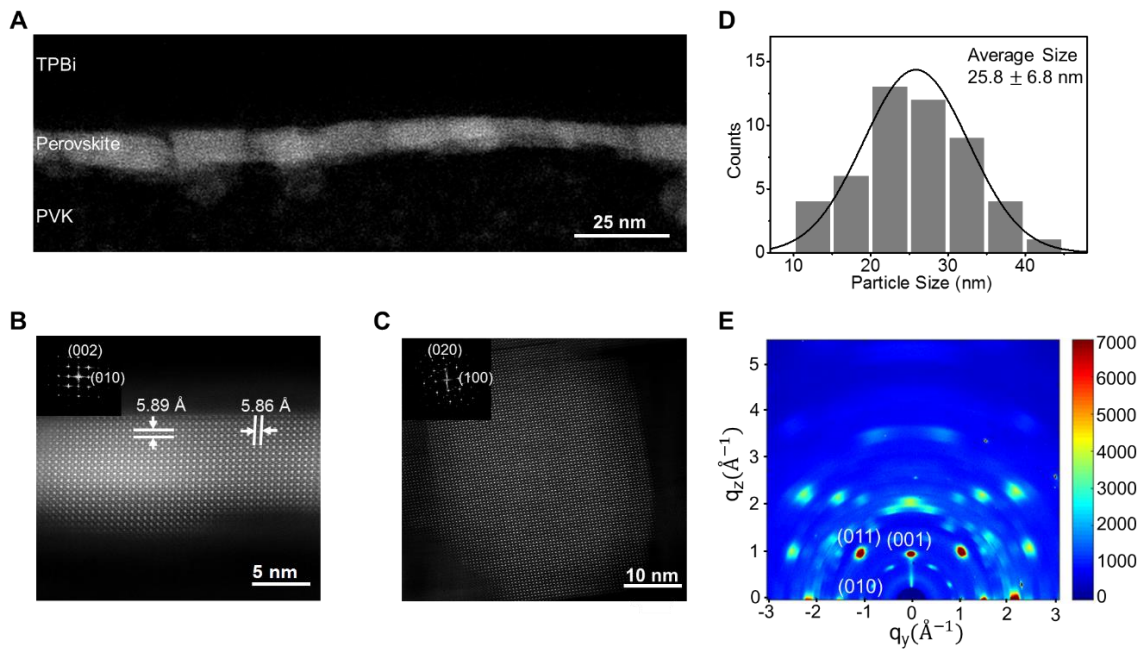

**Fig. 1. Structural characterizations of the perovskite nanoplatelet films.** (**A**) A cross-sectional STEM-HAADF image showing the continuous and pinhole-free perovskite layer. (**B**) A zoomed-in STEM-HAADF image showing the fine structure of a perovskite nanoplatelet. Inset, the corresponding fast Fourier transform (FFT) pattern. (**C**) A typical HRTEM image of the perovskite nanoplatelets dispersed on a copper grid. Inset, the corresponding FFT pattern. (**D**) Statistical diagram of the size distribution of the nanoplatelets measured by HRTEM. The average size is 25.8 nm and the corresponding standard deviation is 6.8 nm. The Gaussian fitting is provided as a guide to the eye. (**E**) GIWAXS pattern. The diffraction spots originate from the crystal faces of nanoplatelets. The two diffraction spots at $q_z$ = 1.065 and $q_y$ = 1.070 Å$^{-1}$ correspond to {001} and {010} of β-CsPbBr$_3$, respectively.

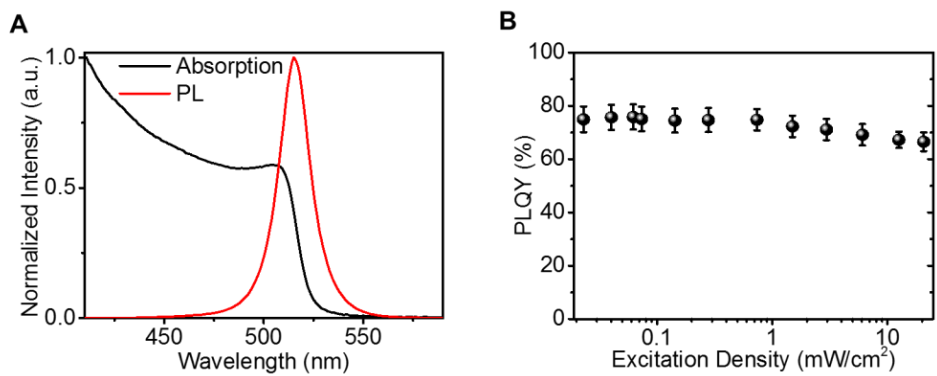

**Fig. 2. Optical properties of the perovskite nanoplatelet films.** (**A**) Absorption and PL (excited by a 405 nm laser) spectra. (**B**) Excitation-intensity-dependent PLQY. The error bars represent the experimental uncertainties in the PLQY measurements at 0.4 mW/cm$^2$ and the errors in the determination of relative PL intensities and excitation power.

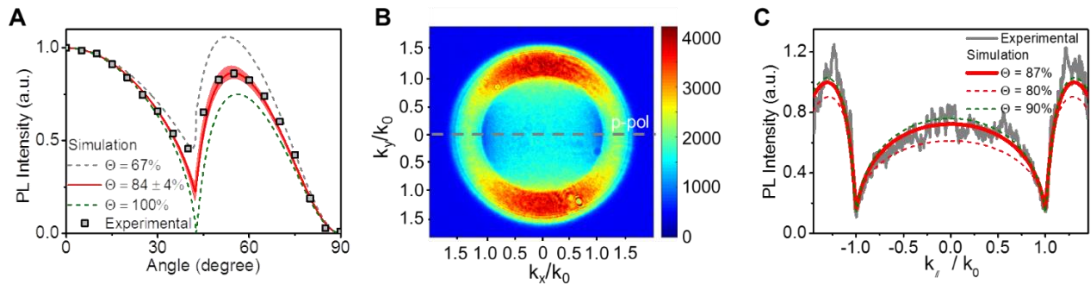

**Fig. 3. Orientations of the TDMs of the perovskite nanoplatelet films.** (**A**) Angle-dependent PL measurements of the perovskite film on a quartz/TFB/PVK substrate. The experimental data (grey squares) are fitted by the classical electromagnetic dipole model (red line), giving a horizontal TDM ratio of 84 ± 4%. (**B**) BFP image of a perovskite film. (**C**) A p-polarized line cut (grey line) along the dashed line in of the BFP image (panel **B**). This line cut is fitted with a horizontal TDM ratio of 87% (red solid line).

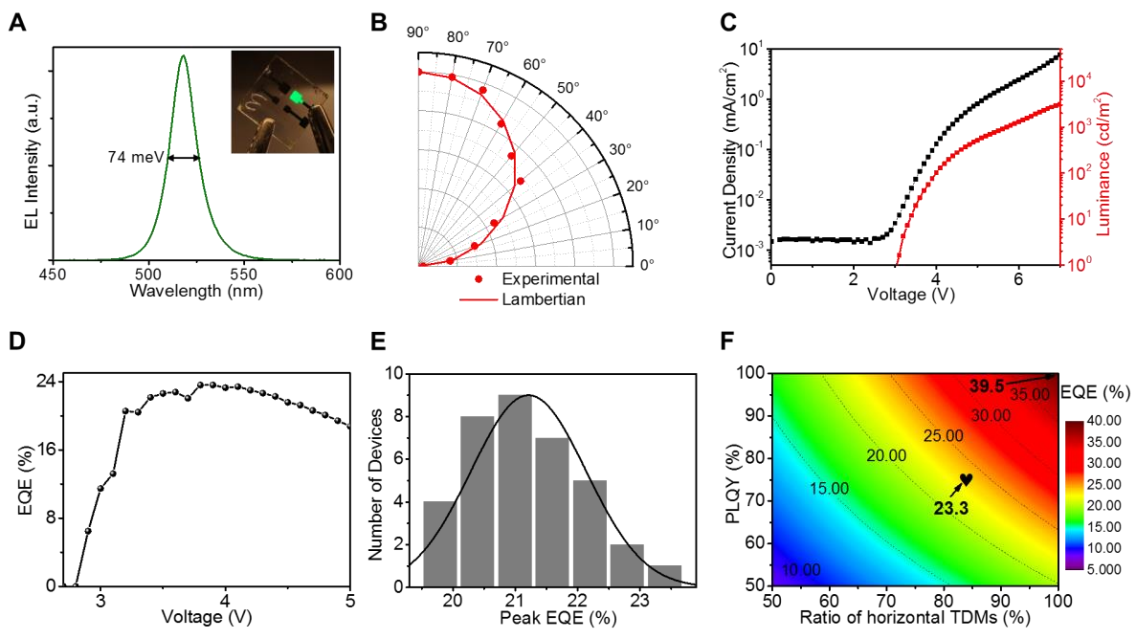

**Fig. 4. Device characterizations of the green LEDs based on the perovskite nanoplatelet films.** (**A**) EL spectrum. Inset, a photograph of an operating green LED (effective area: 3.24 mm$^2$). (**B**) Angular distribution of the EL intensity follows the Lambertian profile. (**C**) Current density-luminance-voltage characteristics of a typical device. (**D**) EQE-voltage relationship of the device with a champion EQE of 23.6%. (**E**) A histogram of peak EQEs from 36 devices. The Gaussian fits are provided as a guide to the eye. (**F**) Contour plot of the simulation results of device EQE as a function of PLQY and Θ of the perovskite emissive layer. The device structure shown in Fig. 4A is used for the simulation. The refractive indexes of the multilayers are obtained by ellipsometer. For our perovskite nanoplatelet film with a PLQY of ~75% and a Θ of 84%, the optical simulation predicts a maximum EQE of ~23.3%.

# Supplementary Materials for

## Efficient light-emitting diodes based on oriented perovskite nanoplatelets

Jieyuan Cui, Yang Liu, Yunzhou Deng, Chen Lin, Zhishan Fang, Chensheng Xiang, Peng Bai, Kai Du, Xiaobing Zuo, Kaichuan Wen, Shaolong Gong, Haiping He, Zhizhen Ye*, Yunan Gao, He Tian, Baodan Zhao, Jianpu Wang and Yizheng Jin*

*Corresponding authors. Email: yezz@zju.edu.cn; yizhengjin@zju.edu.cn

**This PDF file includes:**

Fig. S1. Schematic diagram of three orthogonally-oriented TDM components in planar LEDs.

Fig. S2. Simulated mode distributions inside a planar LED.

Fig. S3. Statistical diagram of the thicknesses of the perovskite film measured from the cross-sectional STEM images.

Fig. S4. Atom force microscopy (AFM) height image of a perovskite nanoplatelet film.

Fig. S5. Additional HRTEM images showing the lateral sizes of the perovskite nanoplatelets.

Fig. S6. Schematic diagram of the setup for the measurement of angular-dependent photoluminescence.

Fig. S7. A schematic illustration of the BFP measurement.

Fig. S8. P-polarized line cut of the BFP images of 4 spots from different regions.

Fig. S9. Perovskite films prepared with precursor solutions with different concentrations of PBABr and PEABr.

Fig. S10. BFP results of the perovskite films processed from the precursor solution with different concentrations of PBABr and PEABr.

Fig. S11. BFP characterizations of perovskites deposited with different (A) concentrations of precursor solutions; (B) choices of bulky organic ligands and (C) choices of substrates.

Fig. S12. Perovskite nanoplatelet films processed from the precursor solution without LiBr.

Fig. S13. Excitation-intensity-dependent PLQY of the perovskite film processed the precursor solution without LiBr.

Fig. S14. Optical properties of $CsPbBr_3$ prepared with precursor solutions with or without the introduction of LiBr.

Fig. S15. CIE colour coordinates of our perovskite LEDs (star).

Fig. S16. EL spectra of our device at different viewing angles.

Fig. S17. Operation stability of device based on the oriented perovskite nanoplatelets.

Fig. S18. Optical simulation of the perovskite LEDs.

Table S1. Comparison of the EQE and FWHM of our perovskite nanoplatelet LED with those of other high-efficiency perovskite LEDs.

Table S2. Comparison of our perovskite nanoplatelet LED with other solution-processed planar LEDs based on colloidal nanostructures with anisotropic optical properties.

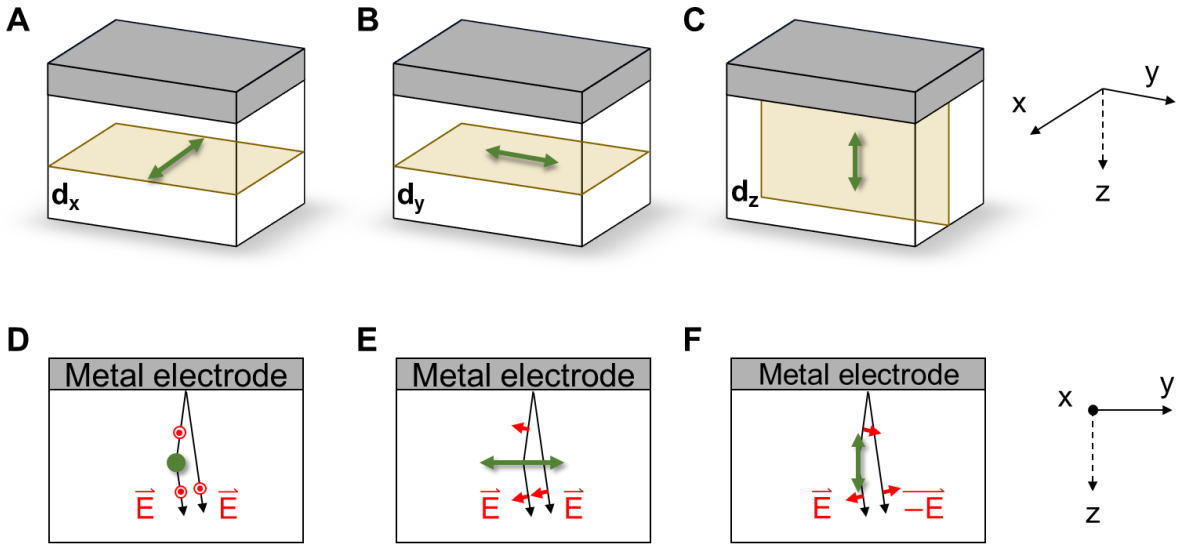

**Fig. S1. Schematic diagram of three orthogonally-oriented TDM components in planar LEDs.** (**A - B**) Two horizontally-oriented TDM components with respect to the electrode surface (green arrows). (**C**) the vertically-oriented TDM component. (**D - F**) Light interference conditions for the three TDM components in the microcavity where the directions of electric fields of the light are denoted by red arrows. For the horizontal TDMs, the interference conditions between the forward-emitted light and the reflected light are constructive, while the vertical TDM experiences the deconstructive interference condition. This fact suggests that optical processes for the vertical dipole in the microcavity are substantially different from those of the horizontal dipoles.

As shown in the above coordination system, the polarization characteristics of radiation from **d$_x$**, **d$_y$** and **d$_z$** are s-polarized, p-polarized, and p-polarized, respectively. The total radiated power (P) is obtained according to the following equation:

$$P = \Theta P_\parallel + (1 - \Theta)P_\perp = \Theta(P_x^s + P_y^p) + (1 - \Theta)P_z^p$$

where Θ is the horizontal dipole ratio defining the orientation of the TDMs. Polarization characteristics are denoted by superscripts. For isotropic emitters that possess randomly oriented dipoles with equal probability for all directions, Θ equals 2/3.

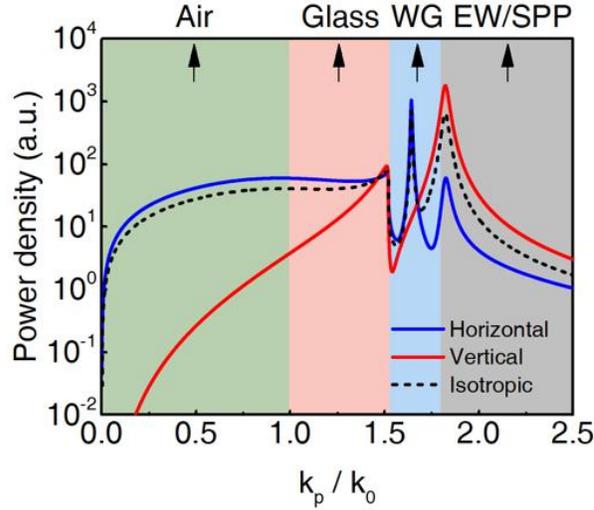

**Fig. S2. Simulated mode distributions inside a planar LED.** In the emissive zone of a planar LED, optical modes can be classified into different regimes according to the ratio of the in-plane wavevector ($k_p$) to the wavevector in vacuum ($k_0 = \frac{2\pi}{\lambda}$). The modes with $0 < \frac{k_p}{k_0} < 1$ will be outcoupled into the air ("Air" modes, green-shaded region). The modes with $1 < \frac{k_p}{k_0} < n_{glass}$, where $n_{glass}$ is the refractive index of the glass substrate (1.52), are trapped in the glass because of the total internal reflection at the glass/air interface ("Glass" modes, red-shaded region). The modes with $n_{glass} < \frac{k_p}{k_0} < n_{emi}$, where $n_{emi}$ is the refractive index of the emissive layer, are trapped inside the device as waveguided modes ("WG" mode, blue-shaded region). In the regime of $\frac{k_p}{k_0} > n_{emi}$, the evanescent waves are coupled to surface plasmon modes at the metal electrode interface via the near-field interactions ("EW/SPP" modes, grey-shaded region). The simulation results show that light emitted by the horizontal TDM (blue line) is readily outcoupled into the air. In contrast, for vertical TDM (red line), most of the light is coupled to surface plasmon modes.

We simulate radiation characteristics of **d$_x$**, **d$_y$** and **d$_z$** in the planar device independently to obtain the mode distributions. The power coupled to different optical modes, which are defined by their in-plane wavevectors in the medium ($k_p$), are separately calculated. The total power radiated from the horizontal dipole component and the vertical dipole component is obtained by integrating power densities of all optical modes:

$$P_{\parallel} = P_x^s + P_y^p = \int_0^\infty F_x^s(k_p)dk_p^2 + \int_0^\infty F_y^p(k_p)dk_p^2$$

$$P_{\perp} = P_z^p = \int_0^\infty F_z^p(k_p)dk_p^2$$

where $F_x^s$, $F_y^p$ and $F_z^p$ are power densities per unit $k_p^2$ radiated from $\mathbf{d_x}$, $\mathbf{d_y}$ and $\mathbf{d_z}$, respectively. In brief, the determination of $F_x^s$, $F_y^p$ and $F_z^p$ includes the calculation of complex Fresnel coefficients on both sides of the emissive layer and the calculation of multiple interferences in the microcavity, which requires complex refractive indices and thicknesses of all materials in the LED as input parameters. These parameters are obtained by ellipsometer measurements.

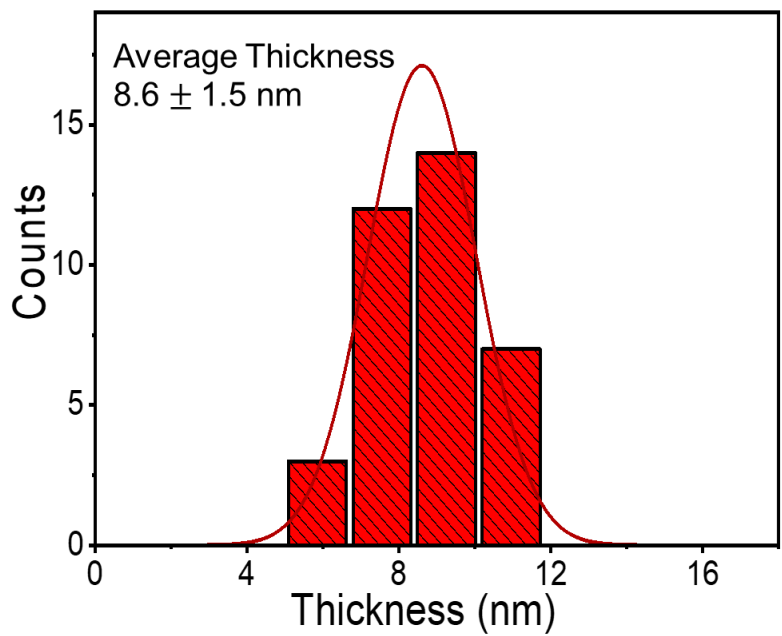

**Fig. S3. Statistical diagram of the thicknesses of the perovskite film measured from the cross-sectional STEM images.** The average thickness and the corresponding standard deviation are determined to be 8.6 nm and 1.5 nm, respectively. The Gaussian fitting is provided as a guide to the eye.

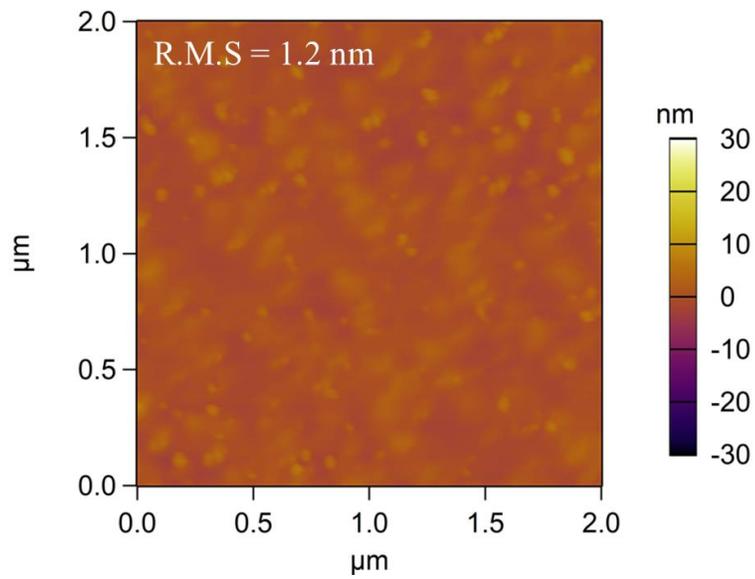

**Fig. S4. Atom force microscopy (AFM) height image of a perovskite nanoplatelet film.** The root-mean-square (RMS) surface roughness is determined to be ~1.2 nm.

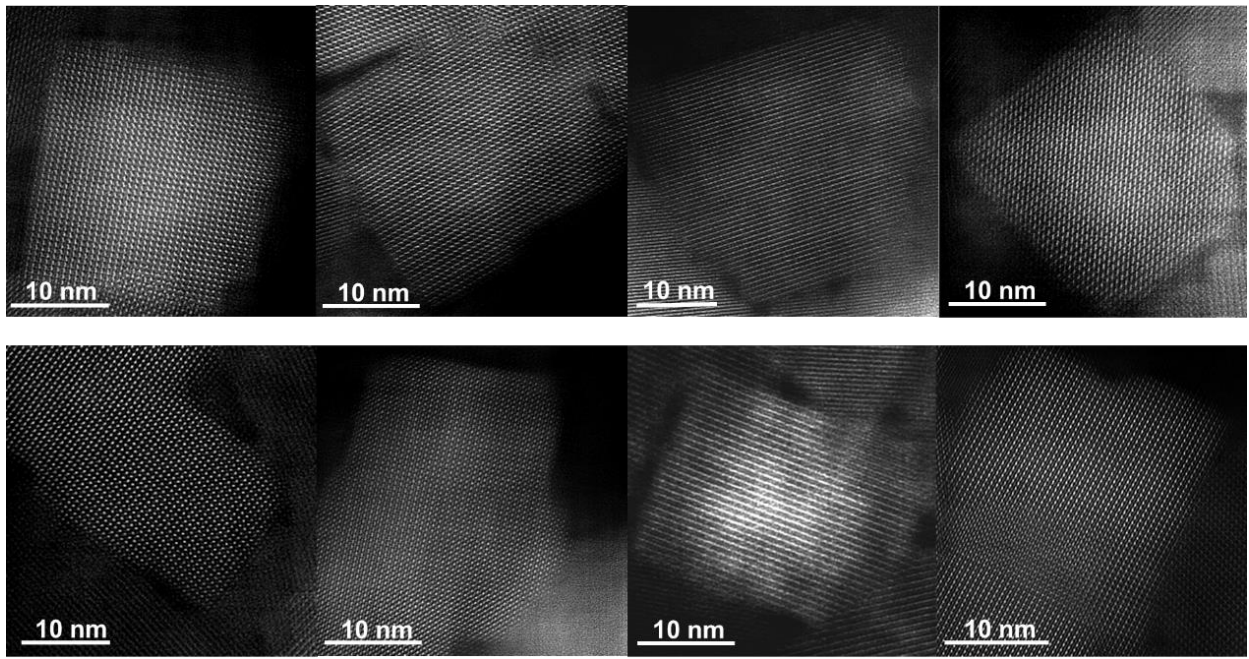

**Fig. S5. Additional HRTEM images showing the lateral sizes of the perovskite nanoplatelets.**

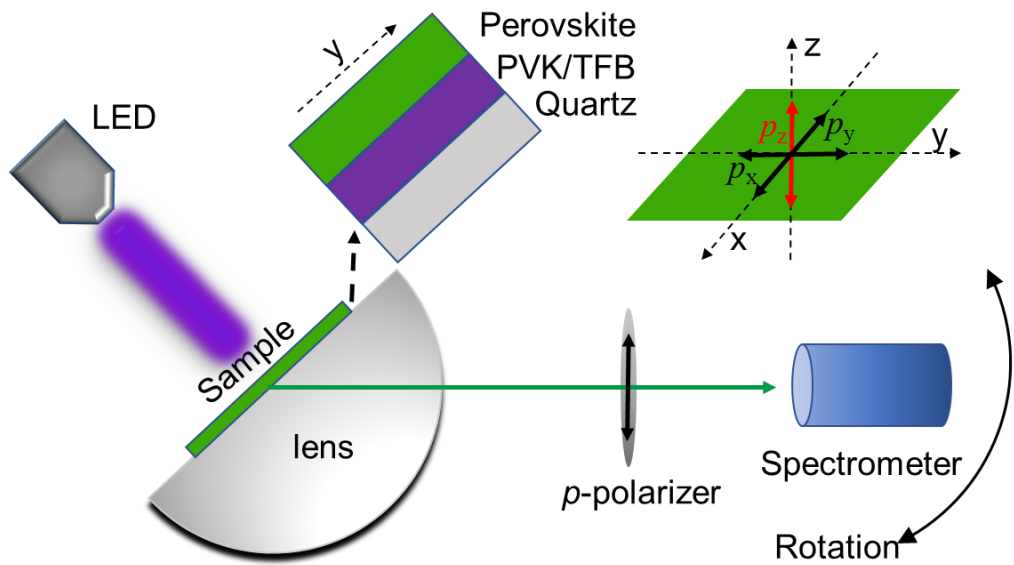

**Fig. S6. Schematic diagram of the setup for the measurement of angular-dependent photoluminescence.** The setup consists of a light source (365 nm LED), a p-polarizer, a spectrometer, and a silica half-prism. The samples are adhered to the silica cylindrical lens, which can effectively break the total-internal-reflection conditions and extracts all the light inside the quartz substrate. Thus, Fig. 3A manifests the intensity distributions inside the substrate. The p-polarizer and the spectrometer are fixed on a rotation stage to collect PL emission at different angles. The dipoles in the emitters are composed of $p_x$, $p_y$, and $p_z$ dipoles according to the orientations, where $p_x$ and $p_y$ dipoles are parallel to the substrate (defined as horizontal dipoles, $p_{/\!/}$) and $p_z$ dipoles are perpendicular to the substrate (defined as vertical dipoles, $p_\perp$). The s-polarized light solely shows information from horizontal dipoles while the p-polarized light arises from both the horizontal dipoles and the vertical dipoles. Thus, the orientation of dipoles is analyzed by the p-polarized light which contains information of both $p_\perp$ and $p_{/\!/}$.

The angular-dependent PL intensity, i.e., the emission pattern of perovskite films in Fig. 3A is affected by both the photonic environment of TDMs and the intrinsic orientation of TDMs. We simulate emission patterns of perovskite films using the classical electromagnetic model described in Fig. S2, except that the top surface of the perovskite film is exposed to the air (n = 1). Providing the thicknesses and complex refractive indices of the multi-layer structure depicted in Fig. 1A, we can calculate power density distribution in the k-space corresponding to arbitrary Θ. The angular-dependent power density inside the quartz substrate ($P(\alpha_{out})$) is derived by the following equation:

$$P(\alpha_{out}) = \frac{k_{pe}^2 \cos \alpha_{out}}{\pi} T_{sub}(k_p)$$

where $\alpha_{out}$, $T_{sub}$, $k_{pe}$ and $k_p$ are the exit angle of light mode transmitted into the substrate, the corresponding transmitted power density in k-space, the wavevector inside the perovskite film and the in-plane wavevector of the light mode, respectively. The ratio of horizontal TDMs of the sample is determined by comparing the experimental results with emission patterns simulated with various Θ.

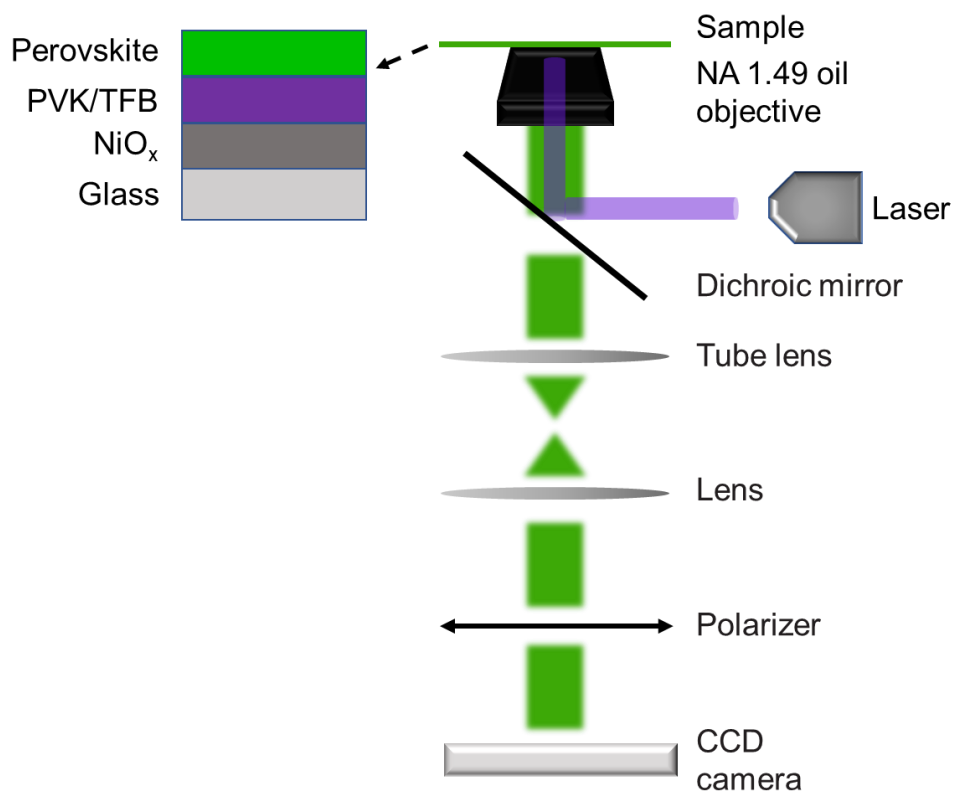

**Fig. S7. A schematic illustration of the BFP measurement.** The sample (NiO$_x$/ TFB/ PVK/ perovskite deposited on a glass slide) was placed on the focal plane of the objective lens and excited with a laser (wavelength: 457 nm; spot size: 750 nm). A high numerical aperture microscope objective (CFI Apo TIRF 60XC Oil, NA 1.49) was used to collect the angular distribution of PL emission from the sample. Polarized BFP images were collected by a CCD camera. Each point in the BFP pattern corresponds to a different angle of emission, which is defined by the photon momentum k and the orientation of the polarizer placed in front of the imaging array.

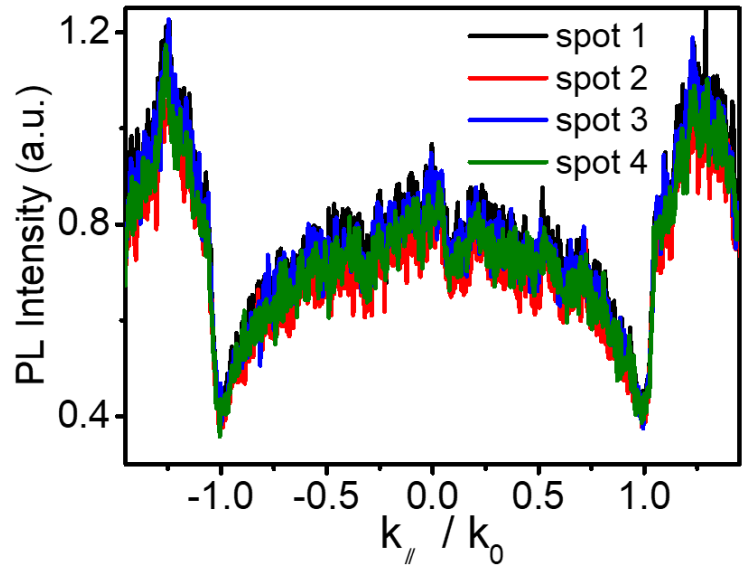

**Fig. S8. P-polarized line cut of the BFP images of 4 spots from different regions.**

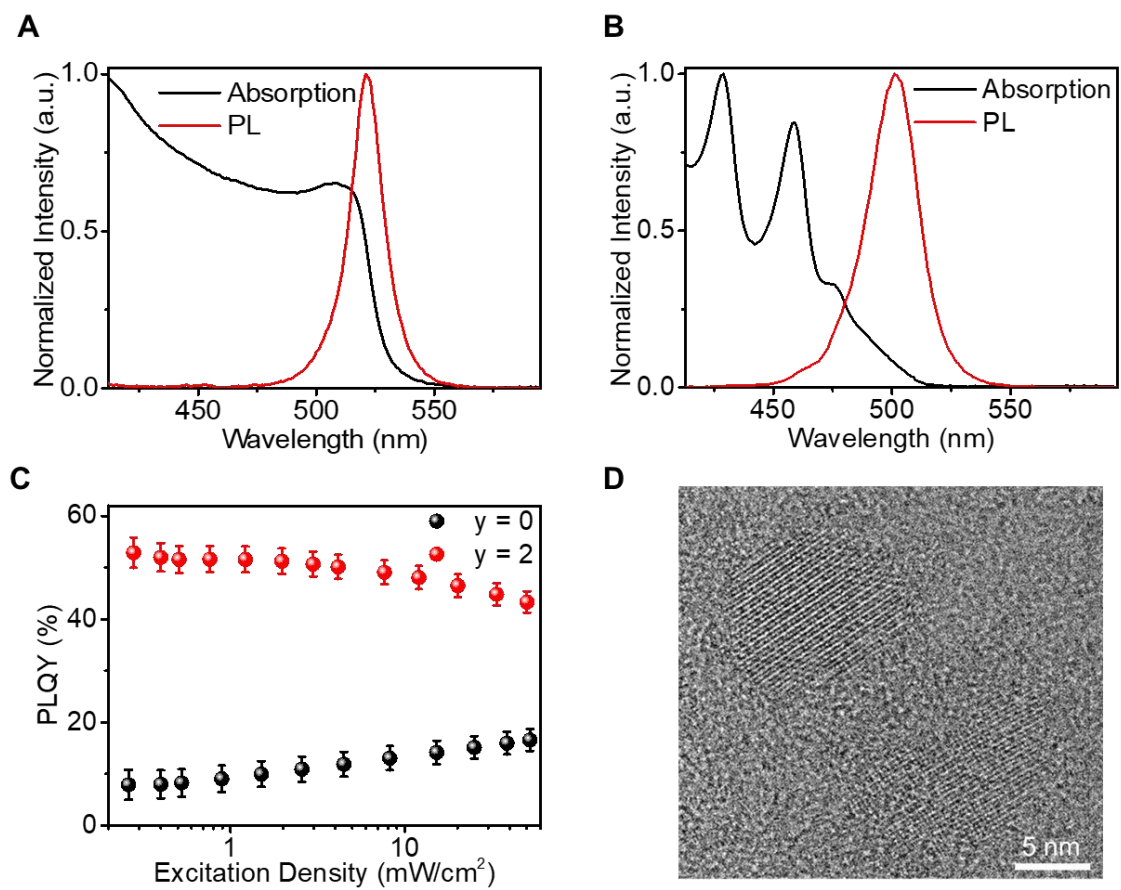

**Fig. S9. Perovskite films processed from the precursor solution with different concentrations of PBABr and PEABr.** The compositions of the precursor solution are abbreviated as $(LiBr)_{0.25}(PBABr)_{3y/4}(PEABr)_{y/4}(CsBr)_{1.75}(PbBr_2)_{1.4}$, where y represents the total amount of the bulky organic ammonium cations. (**A**) Absorption and PL spectra (excitation wavelength: 405 nm) of the y = 0 perovskite film. (**B**) Absorption and PL spectra of the y = 2 perovskite film. (**C**) Excitation-intensity-dependent PLQY of the y = 0 and y = 2 perovskite films. The error bars represent the experimental uncertainties in the PLQY measurements at 0.4 mW/cm$^2$ and the errors in the determination of relative PL intensities and excitation power. (**D**) A typical HRTEM image of the nanocrystals from the y = 2 perovskites. The results indicate that the emission centers in the y = 2 film are perovskite nanocrystals (~10 nm) rather than nanoplatelets.

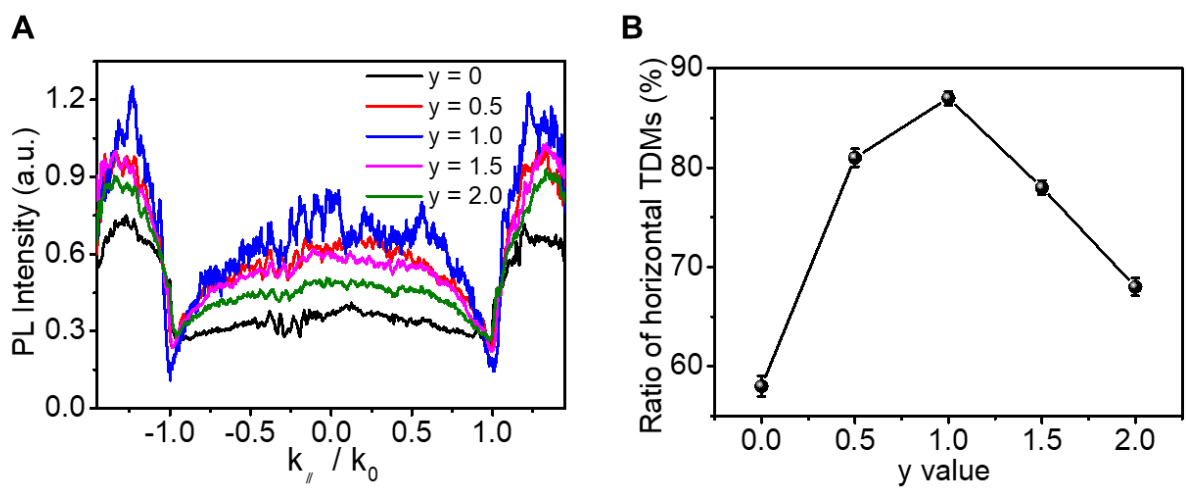

**Fig. S10. BFP results of the perovskite films processed from precursor solution with different concentrations of PBABr and PEABr.** The compositions of the precursor solution are abbreviated as $(LiBr)_{0.25}(PBABr)_{3y/4}(PEABr)_{y/4}(CsBr)_{1.75}(PbBr_2)_{1.4}$, where y represents the total amount of the bulky organic ammonium cations. (**A**) Experimental p-polarized line cut of perovskite films with y = 0, 0.5, 1, 1.5 and 2. (**B**) Ratios of horizontal TDMs determined by the BFP results.

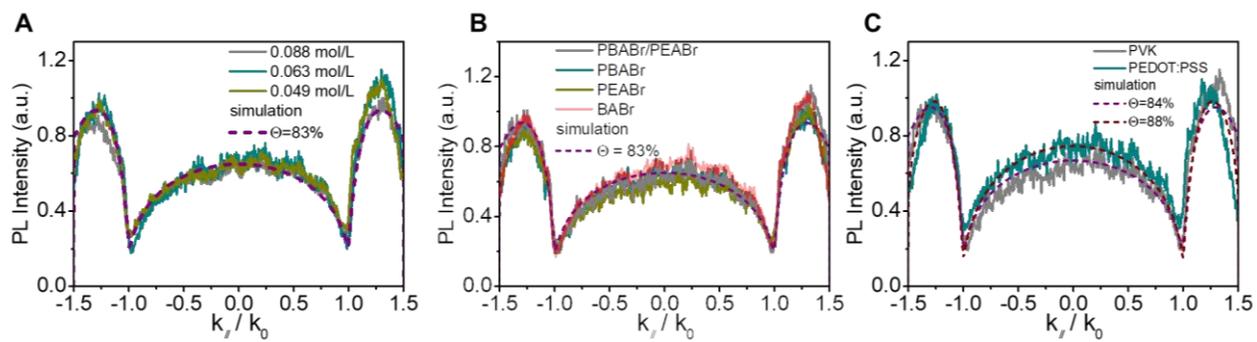

**Fig. S11.** BFP characterizations of perovskites deposited with different **(A)** concentrations of precursor solutions; **(B)** choices of bulky organic ligands and **(C)** choices of substrates.

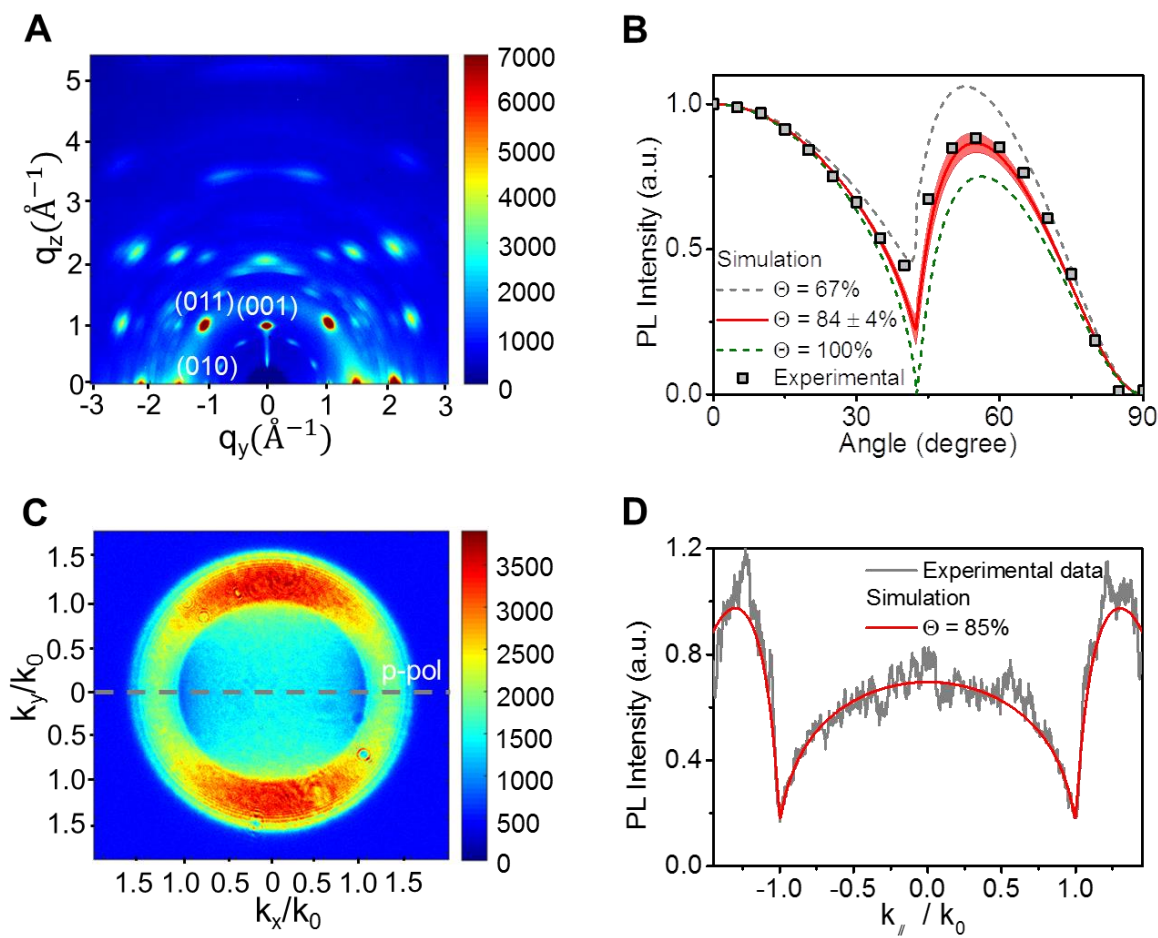

**Fig. S12. Perovskite nanoplatelet films processed from the precursor solution without LiBr.** (**A**) GIWAXS pattern showing that the crystal structure and the orientation of the perovskite crystallites are almost identical to those of the nanoplatelet film processed from the precursor solution with LiBr. (**B**) Angle-dependent PL spectrum. By comparing the results with simulated emission patterns (grey, red and green lines), the ratio of horizontal TDMs is determined to be ~84%, which is same as the film processed from the precursor solution with LiBr. (**C**) BFP image. (**D**) A p-polarized line cut (grey line) along the dashed line in the BFP image (panel **C**). The data is fitted with a ratio of horizontal TDMs of 85% (red solid lines).

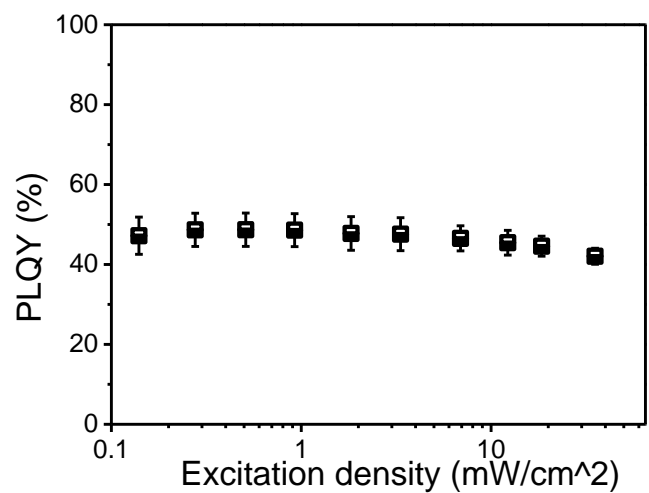

**Fig. S13. Excitation-intensity-dependent PLQY of the perovskite film processed the precursor solution without LiBr.** The error bars represent the experimental uncertainties in the PLQY measurements at 0.4 mW/cm$^2$ and the errors in the determination of relative PL intensities and excitation power.

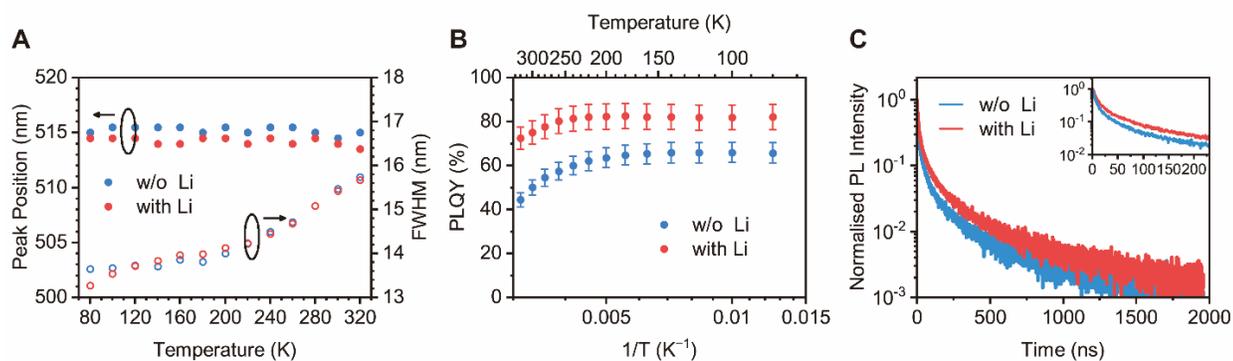

**Fig. S14. Optical properties of CsPbBr$_3$ nanoplatelet films processed from precursor solutions with or without the introduction of LiBr**. (**A**) Temperature-dependent peak position and FWHM (full width at half maximum). (**B**) Temperature-dependent PLQYs. We note that the temperature-dependent PLQYs are calculated based on the PLQY measured at room temperature (around 300 K) and the relative PL intensity at temperature through 80 K to 320 K. The relative standard deviation of PLQY measurements and the fluctuation of absorption with the temperature change is considered in error bars. (**C**) PL decay curves of the films at room temperature. The scope ranging from 0-230 ns is zoomed in in the inset.

Specifically, both samples show identical PL peak positions and similar temperature-dependent tendencies of full-width at half maximum (FWHM) (Fig. S14**A**), indicating their similar exciton binding energies and electron-phonon interactions. These features imply that the two samples should exhibit similar thermal-induced quenching, which stems from exciton disassociation induced by electron-phonon coupling. Furthermore, the PLQYs of the with-Li sample are higher than those of the without-Li sample in the whole temperature range of 80-330 K (Fig. S14**B**). Transient PL results show that the with-Li sample exhibit a longer lifetime of excitation states (Fig. S14**C**). These results indicate a scenario that the with-Li sample possesses fewer energy levels below excitonic levels and thereby, fewer nonradiative losses of photo-generated excitons. We suggest that the introduction of LiBr in the precursor solution may result in perovskite nanoplatelets with better surface passivation.

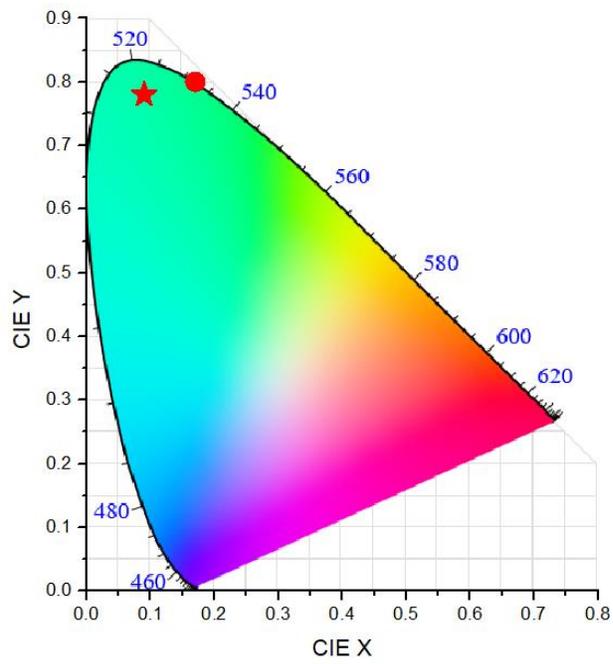

**Fig. S15. CIE colour coordinates of our perovskite LEDs (star).** The red dot represents the coordinates of green Rec 2020 required for wide-colour-gamut displays.

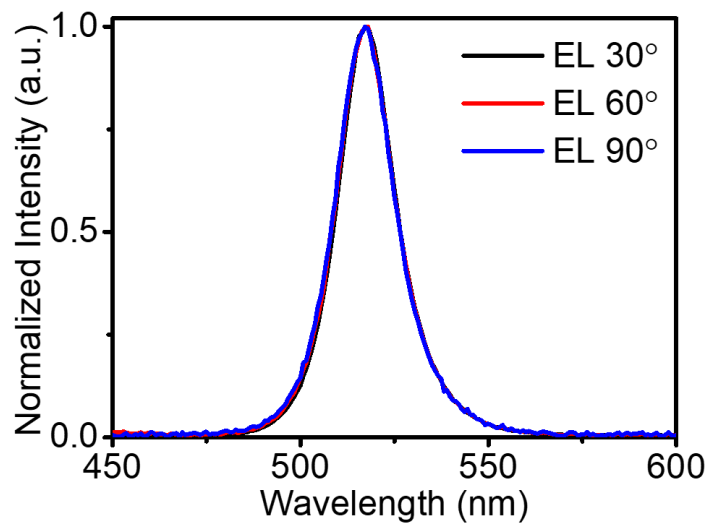

**Fig. S16. EL spectra of our device at different viewing angles.** The EL spectra do not change at different viewing angles.

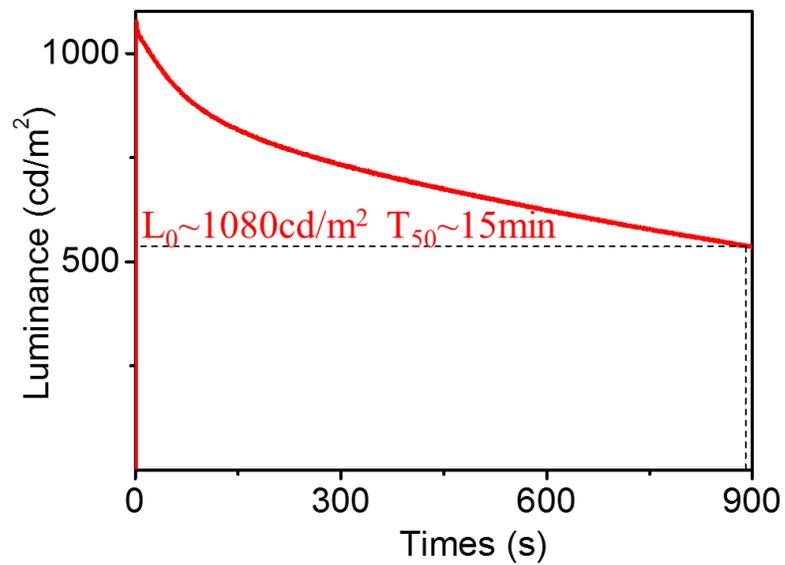

**Fig. S17. Operation stability of device based on the oriented perovskite nanoplatelets.** The device showed a typical $T_{50}$ lifetime of ~15 min at an initial luminance of ~1080 cd/m$^2$).

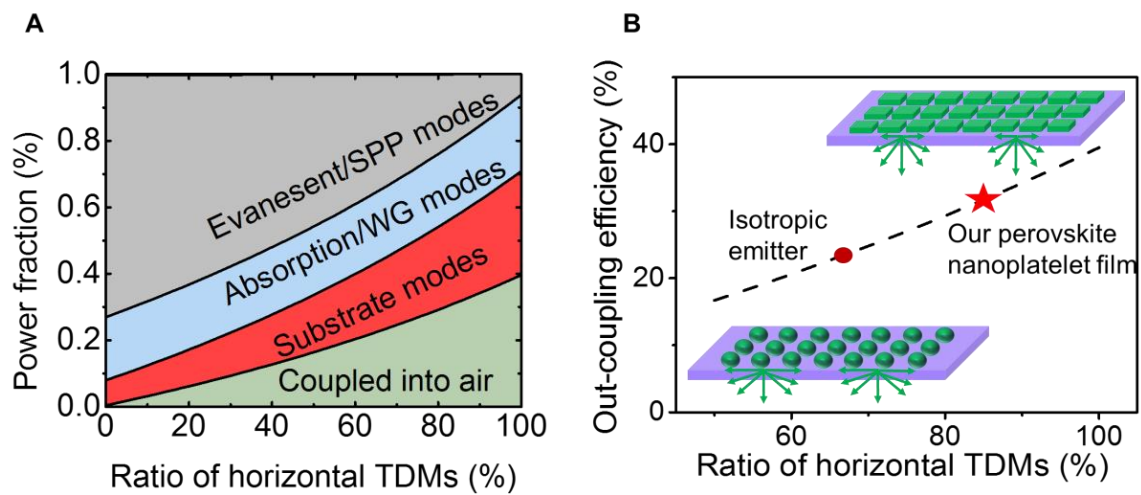

**Fig. S18. Optical simulation of the perovskite LEDs.** (**A**) Simulated power fractions of different optical modes in the perovskite LED as a function of the ratio of horizontal TDMs. The power fraction of outcoupled light (green-shaded) increases at higher ratios of horizontal TDMs, mainly due to the reduced power dissipation in the form of surface plasmon modes (grey-shaded). The device structure described in the main text is used for optical simulation. The refractive index values of various layers are obtained by ellipsometer. (**B**) The relationship between the light outcoupling efficiency of LEDs and the ratio of horizontal TDMs of the perovskite emissive layers extracted from panel **A**.

**Table S1. Comparison of the EQE and FWHM of our perovskite nanoplatelet LED with those of other high-efficiency perovskite LEDs.**

| Emitter | structure | Composition | FWHM (meV) | EQE (%) | Reference |
|---|---|---|---|---|---|
| Isotropic | 3D crystals | FAPbI$_3$ | 75 | 20.7 | *(19)* |
| | | CsPbBr$_3$/MABr | 90 | 20.3 | *(20)* |
| | | (NMA)$_2$(FA)Pb$_2$I$_7$ | 106 | 20.1 | *(21)* |
| | | FAPbBr$_3$ | ~80 | 21.6 | *(23)* |
| | colloidal quantum dots | CsPbBr$_3$/OAM-I | 85 | 21.3 | *(42)* |
| | | CsPbBr$_3$ | 83 | 16.48 | *(50)* |
| | | FA$_{0.9}$GA$_{0.1}$PbBr$_3$ | ~97 | 23.4 | *(26)* |
| | | CsPbBr$_3$ | 83 | 19.1 | *(24)* |
| | | CsPbBr$_3$ | ~75 | 22 | *(25)* |
| **Anisotropic** | nanoplatelets | CsPbBr$_3$ | **74** | **23.6** | **This work** |

**Table S2. Comparison of our perovskite nanoplatelet LED with other solution-processed planar LEDs based on colloidal nanostructures with anisotropic optical properties.**

| Deposition method | Structure | Composition | PLQY (%) | EQE (%) | Reference |
|---|---|---|---|---|---|
| PeLEDs based on colloidal nanostructures | nanoplatelet | FAPbBr$_3$ | 92 (film) | 4.5 | (47) |
| | | - | 94 (film) | ~5 | (51) |
| | | CsPbBr$_3$ | - | 0.2 | (52) |
| | | CsPbBr$_3$ | 27 (film) | 0.55 | (53) |
| | | FAPbBr$_3$ | 92 (film) | 3.04 | (54) |
| | | CsPbBr$_3$ | - | 0.1 | (55) |
| | | MAPbBr$_3$ | 85 (solution) | 0.54 | (46) |
| | nanorod | MAPbX$_3$ | - | - | (56) |
| QLEDs based on colloidal nanostructures | nanoplate | CdSe/CdSe$_{0.8}$Te$_{0.2}$ | 85 (solution) | 3.6 | (57) |
| | | CdSe/CdZnS | 20 (film) | 0.63 | (58) |
| | | CdSe/CdZnS | 40 (film) | 8.39 | (59) |
| | | CdSe/CdS | 60 (solution) | 5.0 | (60) |
| | | CdSe/CdS | 50 (solution) | 0.5 | (61) |
| | | CdS/CdSe/ZnSe | 39 | 1.7 | (62) |
| | | CdSe/CdS | 48 (film) | 5.1 | (63) |
| | nanorod | CdSe/CdS | 54 (film) | 6.1 | (64) |
| | | CdS/CdSe/ZnSe | ~40 (film) | 12.1 | (43) |
| | | CdS/CdSe/ZnSe | - | 10.7 | (65) |
| In-situ grown films | nanoplatelet | CsPbBr$_3$ | 75 (film) | **23.6** | **This work** |